\documentclass[conference]{IEEEtran}

\usepackage{epsfig}
\usepackage{graphicx}
\usepackage{amsmath}
\usepackage{amssymb}
\usepackage{amsxtra}
\usepackage{here}
\usepackage{rawfonts}
\usepackage{times}
\usepackage{url}
\usepackage{cite}
\usepackage{hyperref}
\usepackage{caption}

\usepackage{adjustbox}

\usepackage{epstopdf}

\newtheorem{example}{Example}

\newcommand{\norm}[1]{\left\lVert#1\right\rVert}

\newlength{\aligntop}
\newlength{\alignbot}
\makeatletter

\renewenvironment{align}{%
  \vspace{\aligntop}
  \start@align\@ne\st@rredfalse\m@ne
}{%
  \math@cr \black@\totwidth@
  \egroup
  \ifingather@
    \restorealignstate@
    \egroup
    \nonumber
    \ifnum0=`{\fi\iffalse}\fi
  \else
    $$%
  \fi
  \ignorespacesafterend%
  \vspace{\alignbot}\par\noindent
}

\IEEEoverridecommandlockouts

\begin{document}
\title{\huge How Secure are Multicarrier Communication Systems Against Signal Exploitation Attacks?}
\author{\IEEEauthorblockN{Alireza Nooraiepour, Kenza Hamidouche, Waheed U. Bajwa, Narayan Mandayam} \IEEEauthorblockA{ WINLAB, Department of Electrical and Computer Engineering, Rutgers University, NJ, USA \\
Email: {$\{$kenzaham,narayan$\}$}@winlab.rutgers.edu, {$\{$alireza.nooraiepour,waheed.bajwa$\}$}@rutgers.edu
 \vspace{-0.5cm}}%
\thanks{This work is supported in part by a grant from the U.S. Office of Naval Research (ONR) under grant number N00014-15-1-2168.}
    }
\date{}
\maketitle
\begin{abstract}
In this paper, robustness of non-contiguous orthogonal frequency division multiplexing (NC-OFDM) transmissions is investigated and contrasted to OFDM transmissions for fending off signal exploitation attacks. In contrast to ODFM transmissions, NC-OFDM transmissions take place over a subset of active subcarriers to either avoid incumbent transmissions or for strategic considerations. A point-to-point communication system is considered in this paper in the presence of an adversary (exploiter) that aims to infer transmission parameters (e.g., the subset of active subcarriers and duration of the signal) using a deep neural network (DNN). This method has been proposed since the existing methods for exploitation, which are based on cyclostationary analysis, have been shown to have limited success in NC-OFDM systems. A good estimation of the transmission parameters allows the adversary to transmit spurious data and attack the legitimate receiver. Simulation results show that the DNN can infer the transmit parameters of OFDM signals with very good accuracy. However, NC-OFDM with fully random selection of active subcarriers makes it difficult for the adversary to exploit the waveform and thus for the receiver to be affected by the spurious data. Moreover, the more structured the set of active subcarriers selected by the transmitter is, the easier it is for the adversary to infer the transmission parameters and attack the receiver using a DNN.
\end{abstract}
\section{Introduction}

The explosive growth of Internet of things (IoT) and machine-to-machine (M2M) applications has created many scientific and engineering challenges in cellular systems \cite{cisco,zhou2018effect}. This unprecedented transformation of wireless networks will not only generate a massive amount of traffic, but it will also lead to the emergence of sophisticated cyber-attacks, which will pose a serious threat to security and privacy of communications infrastructure than ever before. The vulnerability of IoT systems is driven by the strong  restrictions on sensors and smart devices in terms of computational capabilities, energy and power. Furthermore, since the devices exchange information via wireless communication links, these will increase the number of false data injections, attacks and information leaks.

In recent years, an extensive literature has focused on the challenges of physical layer security and several techniques have been proposed to deal with jamming and eavesdropping \cite{nichols2001wireless,perrig2004security,wu2013practical}. Spread spectrum \cite{nichols2001wireless} is one of the most widely used techniques to resist jamming by spreading a signal over a wider bandwidth. The authors in \cite{sperandio2002wireless} proposed radio frequency (RF) fingerprinting technology, which consists of extracting RF features to define the RF identification data of a wireless device. This technique generates an alert when an unknown fingerprint is detected, thereby, helping to distinguish an attacker from a legitimate transmitter. The authors in \cite{wu2013practical} proposed two security schemes for the multiple input multiple output (MIMO) systems, where the first technique utilizes the precoding matrix indices as secret keys and the second technique quantizes the channel coefficients directly to generate the secret keys.

In contrast to eavesdropping and jamming, only a few works have studied the impact of false data injection via signal exploitation in wireless systems. In signal exploitation-based attacks, malicious users intercept the signals transmitted by a legitimate transmitter, infer the transmit parameters that are necessary to reconstruct the signal waveform (e.g., total bandwidth, start of a signal and symbol duration) and then use this information to send spurious data and attack the receiver as shown in Fig. \ref{fig:model}. In practical systems, remote attackers could compromise patients' medical devices \cite{BigThink} or autonomous vehicles \cite{Vehicules}, which may not only cause huge economic losses to individuals but also threaten peoples' lives.

The aforementioned physical attacks, in particular signal exploitation attacks, are shown to be able to compromise orthogonal frequency division multiplexing (OFDM) systems, which are widely used in wireless communication networks \cite{nichols2001wireless}. In contrast to ODFM transmission, non-contiguous (NC)-OFDM transmission takes place over a subset of active subcarriers to either avoid incumbent transmissions or evade jamming and eavesdropping attacks. Aside from strategically exploiting the available spectrum, NC-OFDM systems \cite{rajbanshi2006efficient,kumbhkar2015opportunistic} are also shown to make multicarrier systems less vulnerable to false data injection attacks. In this regard, authors in \cite{sridharan2016physical} examined the low probability of exploitation (LPE) characteristics of NC-OFDM transmission assuming that an adversary is using cyclostationary analysis \cite{cycl1} to infer transmission parameters. Cyclostationary analysis, which is based on cyclic auto-correlation function of the received signal, is the main tool used by adversaries to infer transmission parameters in different works \cite{CAF1,CAF2,CAF3} that study security for multicarrier systems. In \cite{sridharan2016physical}, the authors showed that the cyclostationary analysis is extremely challenging to do for most choices of NC-OFDM transmission parameters. Therefore, one can come up with LPE-centric designs using NC-OFDM schemes, assuming cyclostationary analysis is the only tool being used at the adversary. However, it is not clear if this is still the case when the adversary utilizes more powerful machine learning tools for the exploitation attacks.

The main goal of this paper is to take the next natural step and assume that the adversary is equipped with deep learning tools, which have been recently shown to be a promising way for solving analytically intractable problems in communication systems. In particular, the authors in \cite{Deepsig} proposed an end-to-end learning of communication systems based on deep neural networks (DNNs). They optimize transmitter and receiver jointly without considering the classical communication and signal processing blocks including channel encoder and modulator. In \cite{o2017introduction}, the authors showed that deep learning techniques are very promising in scenarios where the channel is too complex to be described analytically. Furthermore, the authors in \cite{overtheair} demonstrated how a DNN-based system can communicate over-the-air without the need for any conventional signal processing block.

The main contribution of this paper is to investigate the LPE characteristics of NC-OFDM systems assuming the presence of an adversary capable of utilizing machine learning tools \cite{o2017introduction}. To this end, we consider a point-to-point communication between a transmitter and a receiver in the presence of an adversary (exploiter). The goal of the exploiter is to infer transmission parameters, reconstitute the waveform, and then send spurious data to the legitimate receiver using the estimated waveform. We assume that the adversary uses DNNs for estimating the transmission parameters. The signal received by the receiver is analyzed under different communication schemes to determine to what extent the adversary can exploit the signal. Simulation results show that an adversary utilized with DNNs is able to exploit a signal which considered to be completely secure against cyclostationary based attacks at the expense of higher complexity (related to training a DNN based on the received data). Specifically, these results show that the adversary can accurately infer the transmission parameters under OFDM transmission. However, for NC-OFDM transmissions, the success of exploitation attacks highly depends on how the active subcarriers are chosen at the transmitter; the more structured is the band allocation, the better is performance of the adversary in estimating the band allocation pattern and exploiting the signal. We assume that the exploiter is able to build a dataset over time based on the received noisy signals and their corresponding true transmission parameters including band allocation pattern and sub-carrier spacing (as lables), and use them for training DNN's. 

The rest of the paper is organized as follows. In section \ref{System Model}, the system model is introduced. An introduction to cyclostationary analysis and its limitations for signal exploitation attacks is presented in Section \ref{Signal exploitation using cyclostationary analysis}. In Section \ref{sec:dlm}, the problem of signal exploitation using DNNs is formulated and the architecture of the neural network is presented. Section \ref{sec:simulations} presents the simulation results. Conclusions are presented in Section \ref{sec:conclusion}.

\section{System Model}\label{System Model}
\begin{figure}
\vspace{-50pt}
\includegraphics[scale=0.3]{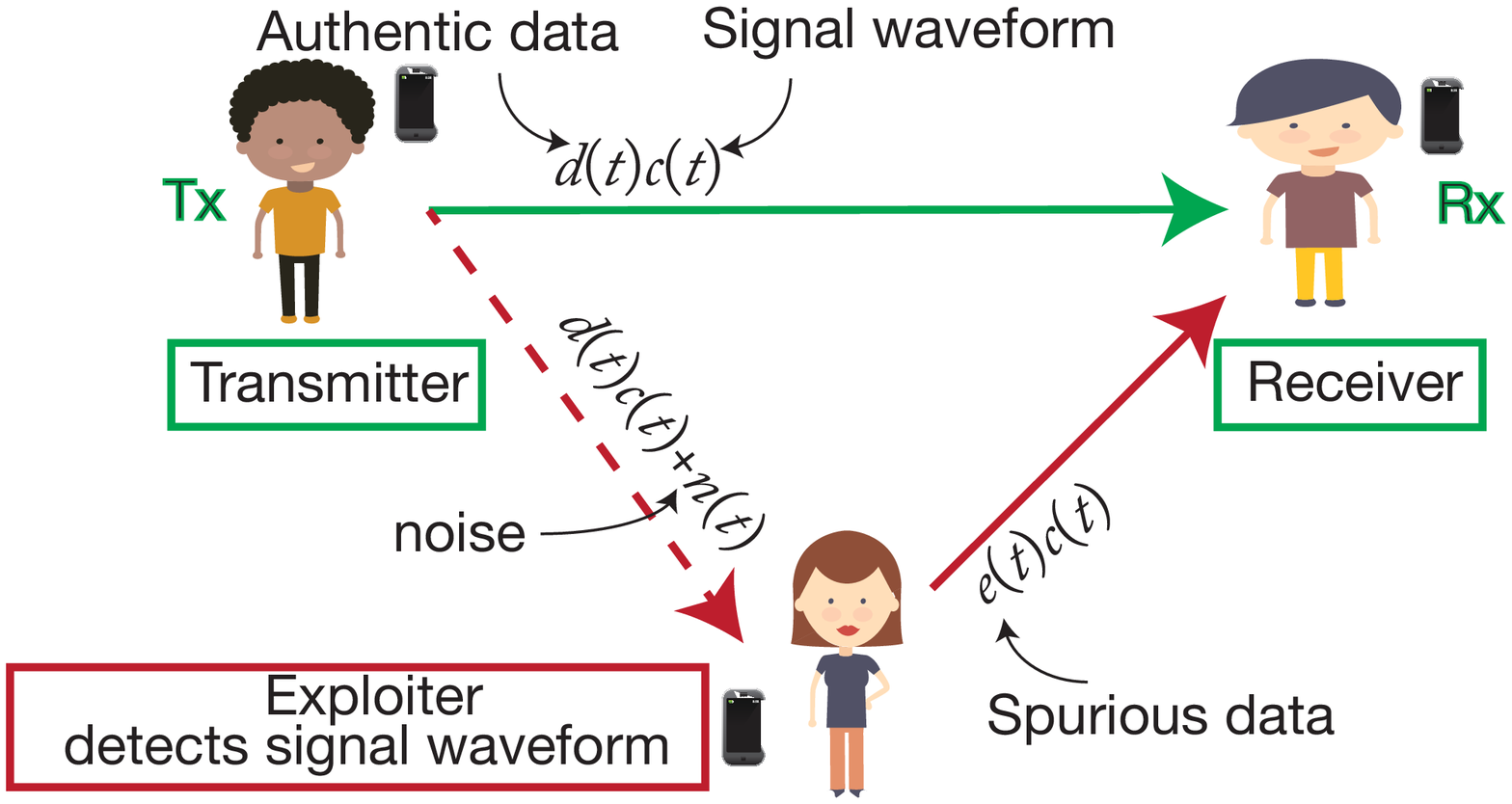}
\caption{A schematic representation of a signal exploitation attack.}
\label{fig:model}
\vspace{-15pt}
\end{figure}

Consider a system composed of a transmitter, receiver and an adversary (exploiter).  The goal of the exploiter is to overhear the signals sent by the transmitter to the legitimate receiver to infer the transmit parameters and use this information to send spurious data to the receiver. The point-to-point communication link between the transmitter and receiver is assumed to operate over a total bandwidth $B$ composed of a set of $N$ subcarriers. The transmitter can either transmit over the whole band when using OFDM transmissions or a subset $\mathcal{S}$, which represents active subcarriers, in the case of NC-OFDM transmission. The transmitted signal can be written as:
\begin{equation}
\label{NCOFDMSignal}
s(t)=\sum_{m=-\infty}^{m=+\infty}\sum_{n\in\mathcal{S}} s_{m,n}e^{j2\pi f_n(t-mT_o)}g(t-mT_o),
\end{equation}
where $s_{m,n}$ is the symbol transmitted on the $n$th subcarrier at the $m$th time slot, $T_o$ is the duration of one NC-OFDM symbol and given by $T_o=T_u+T_{cp}$, with $T_u$ and $T_{cp}$ being the useful symbol duration and the duration of the cyclic prefix. $g(\cdot)$ is a rectangular pulse of width $T_o$, centered at $T_o/2$ and $f_n=n\Delta f$ is the center frequency of each subcarrier, with $\Delta f$ being the width of each subcarrier. The received signal at the exploiter is given by:
\begin{equation}\label{receivedsignal}
r(t)=s(t)+n(t),
\end{equation}
where $n(t)$ is the additive white Gaussian noise. In other words, similar to \cite{sridharan2016physical}, we focus our attention to the case where the exploiter is in the line of sight of the transmitter and keep multi-path channel for future investigation. Consequently, we do not consider transmitting cyclic prefix for the OFDM and NC-OFDM signals, i.e., $T_{cp}=0$. An illustration of the system model is given in Fig. \ref{fig:model}, where the exploiter seeks to find the transmission parameters including $\Delta f=1/T_u$, set $\mathcal{S}$, $T_{cp}$ and the function $g(\cdot)$. In this work, we assume that $T_{cp}$ and $g(\cdot)$ are fixed at the transmitter and known at the exploiter. Having all transmission parameters in hand, the exploiter is able to generate waveforms similar to (\ref{NCOFDMSignal}) by injecting spurious data in place of $s_{m,n}$ and transmitting this data to the receiver. Next, we discuss cyclostationary analysis, which is the classical tool for inferring transmission parameter.
\section{Signal exploitation using cyclostationary analysis}\label{Signal exploitation using cyclostationary analysis}
\begin{figure}
\centering
\includegraphics[width=7.5cm]{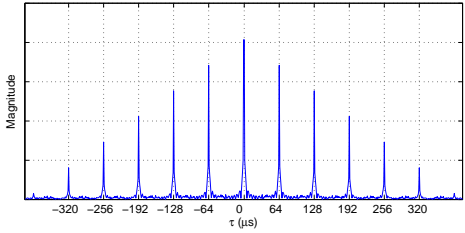}
\caption{Estimated CAF (\ref{estimatedCAF}) at $\alpha=0$ for three cases considered in Table \ref{CAFTable} at SNR $5$ dB where $\tau=\bar{\tau}T_s.$}
\label{fig:CAF}
\vspace{-15pt}
\end{figure}

While cyclic prefix is useful to mitigate the effect of inter-carrier interference in multicarrier systems, it also enables an adversary to infer basic transmission parameters using cyclostationary analysis \cite{CAF1}. Cyclostationary analysis is based on the auto-correlation function $R(t,\tau)$ of the transmitted signal, which can be calculated as
\begin{align}
\label{autocorrelation}
\begin{split}
R(t,\tau)&=E[s(t)s^*(t)]\\&=\sigma^2_s\big(\sum_{n\in \mathcal{S}}e^{j2\pi f_n\tau}\big)\sum_m g(t-mT_0)g^*(t-mT_0-\tau).
\end{split}
\end{align}
The periodicity of $R(t,\tau)$ in $t$ allows representing it as a Fourier series sum
 \begin{align}
\label{autocorrelation2}
\begin{split}
R(t,\tau)&=\sum_n R(\alpha_n,\tau)e^{j2\pi\alpha_nt},
\end{split}
\end{align}
where $\alpha_n=n/T_0$ is the cyclic frequency and $R(\alpha_n,\tau)$ is called the cyclic auto-correlation function (CAF). For OFDM transmissions, this function can provide an exploiter with $T_u$ and $T_0$ \cite{CAF1,CAF2}. However, for NC-OFDM transmissions this analysis does not always lead to the correct results. Specifically, we represent the following example from \cite{sridharan2016physical}.
\begin{example}
\label{CAFexample}
Consider an NC-OFDM signal with a total number of subcarriers $N=256$ where active ones are spaced $q$ subcarriers apart (known as interleaved subcarriers). The transmitter chooses two transmission parameters $q$ and $T_u$ based on one of the cases presented in Table \ref{CAFTable} and sends (\ref{NCOFDMSignal}) over the channel. Then, we assume an exploiter receives the signal (\ref{receivedsignal}), sample it at a rate $1/T_s$ to obtain $M$ samples and estimates the CAF function using
\begin{equation}
\label{estimatedCAF}
\hat{R}(\alpha,\bar{\tau}T_s)=\frac{1}{M}\sum_{n=1}^Mr[n]r^*[n-\bar{\tau}]e^{-j2\pi \alpha n T_s},
\end{equation}
where $\bar{\tau}\in \mathcal{Z}$. To extract $q$ and $T_u$ corresponding to the original transmission, exploiter must look at the locations of the peaks in (\ref{estimatedCAF}) at $\alpha=0$. For the cases in Table \ref{CAFTable}, (\ref{estimatedCAF}) is illustrated in Fig. \ref{fig:CAF}. Specifically, for all three cases, CAF-based analysis results in the same plot. Therefore, exploiter can not decide which one of the transmission parameters' sets are used by the transmitter.
 \hfill $\blacksquare$
\end{example}
\begin{table}[htb] 
\centering
\caption{Three sets of transmission parameters}
\label{CAFTable}
\begin{tabular}{|l|l|l|l|}
\hline
Case & $T_u (\mu s)$   & $T_0(\mu s)$ & $q$ \\ \hline
$1$ & $320$ & $320\leq T_0 \leq 640$  & $5$ \\ \hline
$2$ & $256$ &  $320\leq T_0 \leq 512$ & $4$ \\ \hline
$3$ & $192$ &  $320\leq T_0 \leq 384$ & $3$ \\ \hline
\end{tabular}
\end{table}
The exploiter's situation in example \ref{CAFexample} could get worse as the band allocation becomes more complicated (e.g. when multiple interleaving factors $q_i$, $i=1,2,\dots$ are used in different parts of the band) for which CAF analysis becomes very challenging. In the next section, we study how deep learning can be used to enhance solving problems of this sort. Later on, we will consider Example \ref{CAFexample} again in Section \ref{Comparison to CAF-based analysis} and show how it can be solved using deep learning tools and assuming that the exploiter has access to a reasonable size of training data.
\section{Deep Learning Model for signal exploitation}
\label{sec:dlm}
We now introduce a deep learning approach to signal exploitation attacks in NC-OFDM systems. As discussed in the previous section, our main motivation comes from the fact that previous techniques like cyclostationary analysis \cite{cycl1} for signal exploitation attacks have been shown to be extremely challenging for most choices of transmission parameters \cite{sridharan2016physical}. In contrast, we investigate performance of an exploiter that applies machine learning tools to exploit the signal and attempt to answer the question that if/when she is able to do so and what are the parameters that affect her performance. Next, we first present a DNN-based model for signal exploitation. Then, we consider the problem in Example \ref{CAFexample} and describe how it can be solved using DNNs.

\subsection{DNN-based exploitation}\label{DNN Training}
\begin{figure}
\centering
\includegraphics[scale=0.6]{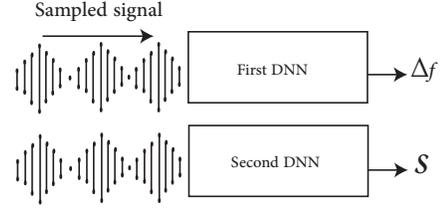}
\caption{Block diagrams of two DNNs used for estimating transmission parameters at the exploiter.}
\label{fig:learning}
\vspace{-15pt}
\end{figure}

We assume the exploiter has access to deep neural networks (DNNs) of fixed input size $n$, which try to infer the transmission parameters introduced in Section \ref{System Model}. Fig. \ref{fig:learning} illustrates this where two separate DNNs are used to infer two different parameters, set $\mathcal{S}$ and $\Delta f$. The exploiter overhears the transmissions between the transmitter and the receiver. Then she collects samples and the corresponding true transmission parameters as the training data. Specifically, we assume that the exploiter samples the received noisy signal in (\ref{receivedsignal}) to extract $n/2$ (complex) samples $r(kT)$ and feed the DNNs with the corresponding real and imaginary parts, i.e.,
\begin{equation}
\label{sampledreceived}
r(kT)=\sum_{n\in\mathcal{S}} s_{m,n}e^{j2\pi f_n(kT-t_0)}+n(kT), k=0,\dots,n_1/2-1,
\end{equation}
where $T=\frac{1}{\Delta fn_1}$ and $t_0$ represents the start of the sampling time. We note that $1/\Delta f$ denotes the duration of the NC-OFDM signal and $1/T$ represents the sampling rate here. We assume the transmitter sends signals with different values of $\Delta f$ and uses various band allocations $\mathcal{S}$ as subsequently explained. 
The specifications for the first DNN in Fig. \ref{fig:learning} are as follows. 
\begin{itemize}
\item {\bf Input: } The real part $\mathcal R(\cdot)$ and the imaginary part $\mathcal I(\cdot)$ of the samples $r(kT)$, $k=0,\dots,n_1/2-1$ in (\ref{sampledreceived}); specifically, the input vector is $[\mathcal R(r(0)), \mathcal R(r(T)), \dots, \mathcal R(r((n_1/2-1)T)), \mathcal I(r(0)), \mathcal I(r(T)), \dots, \mathcal I(r((n_1/2-1)T))]$. We also have fixed $n=192$ throughout this work.
\item {\bf Output: } Estimated subcarrier width $\widehat{\Delta f}$.
\item{\bf Architecture:}  There are $4$ fully connected layers, two of which are hidden layers. The numbers of neurons in the layers are $192$, $100$, $50$ and $1$, respectively. The activation function for all the layers is chosen to be ReLU function. Note that we also tried higher number of hidden layers and higher number of neurons for each layer. However, the performance did not improve in a noticeable way.  
\item{\bf Training:} The weights in each layer of the neural network are determined by minimizing the $L_2$ loss
$(\widehat{\Delta f}-\Delta f)^2$ where $\Delta f$ is the true parameter assumed to be known at the exploiter during the training stage. We have used Tensorflow and Adam optimizer \cite{Adam} with a learning rate of $0.0005$ to train the model accordingly.
\end{itemize}

The second DNN is meant to infer the subcarrier pattern. Its properties are described as follows:
\begin{itemize}
\item {\bf Input:} The same input vector described above for the first DNN.
\item {\bf Output:} The subcarriers' allocation pattern estimation $\widehat{\mathbf{\mathcal{S}}}$, which is a binary vector of fixed size $L=64$, where $0$ and $1$ represent inactive and active subcarrier, respectively.
\item{\bf Architecture:}  There are $6$ fully connected layers, four of which are hidden layers. The number of neurons in the layers are $192$, $350$, $600$, $400$, $200$ and $64$, respectively. The activation function for all the layers is chosen to be ReLU function except for the last layer, where the sigmoid function, $f(x)=1/(1+e^{-x})$, is used to ensure an output in the range $[0,1]$. To see whether a subcarrier is inactive or active, we assume the exploiter further convert these values to $0$'s and $1$'s using the following function
\begin{equation}
g(x)=\begin{cases}
    0,& 0\leq x\leq 0.5,\\
    1,              & 0.5<x\leq 1.
\end{cases}
\end{equation}
\item{\bf Training:} The DNN is trained by minimizing the following distance metric:
\begin{equation}
\label{DNN2Diff}
Dist(\mathbf{\mathcal{S}},\widehat{\mathbf{\mathcal{S}}})=\frac{1}{M}\sum_{i=1}^{L} (\mathbf{\mathcal{S}}(i)-\widehat{\mathbf{\mathcal{S}}}(i))^2,
\end{equation}
where $\mathbf{\mathcal{S}}$ is the true binary allocation pattern assumed to be known at the offline training stage. Each element in $\mathbf{\mathcal{S}}$ is $0$ for inactive subcarriers and $1$ for active ones. We note that if $|\mathbf{\mathcal{S}}|< L$, we append zero's at the end of $\mathbf{\mathcal{S}}$ to make it of length $L$, the same length as the DNN's outputs.
We have used RMSProp optimizer with learning rate $0.001$ in Tensorflow in order to train the second DNN.
\end{itemize}

During the online testing stage, the exploiter receives signals with different transmission parameters, samples them according to (\ref{sampledreceived}) and then feeds the resulting data to the trained model to infer the parameters $\widehat{\mathbf{\mathcal{S}}}$ and $\widehat{\Delta f}$. Finally, she will use these to reconstruct (\ref{NCOFDMSignal}) and send her own data in place of $s_{m,n}$ over the waveform to the receiver.

\begin{figure}[ht!]
\vspace{-10pt}
\centering
\includegraphics[width=8cm]{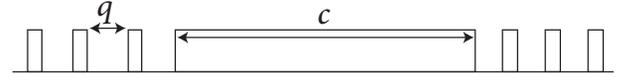}
\caption{NC-OFDM structure $1$ band allocation (struct. $1$). Location of the contiguous bans ($c$) is considered to be random.}
\label{fig:structNC1}
\vspace{-15pt}
\end{figure}

\begin{figure}[ht!]
\centering
\includegraphics[width=8.5cm]{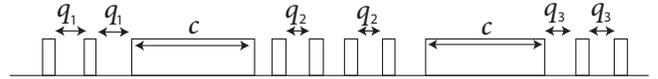}
\caption{NC-OFDM structure $2$ band allocation (struct. $2$). Locations of the contiguous blocks are random.}
\label{fig:structNC2}
\vspace{-10pt}
\end{figure}

Exploiting the transmitter waveforms depends on the structure of the band allocation pattern. We consider four different band allocations in this paper: $1$) A single contiguous block of active subcarriers (OFDM signal). $2$) NC-OFDM signal whose band allocation is illustrated in Fig. \ref{fig:structNC1} where $q$ denotes the number of inactive subcarriers between active ones and $c$ denotes the length of a block of contiguous active subcarriers. We refer to it as \textit{structure $1$} in the following. For training and test purposes, we generate signals of this type with $q$ in range $[1,6]$ and $c$ in range $[4,53]$ where the location of $c$ in the band is considered to be random. $3$) NC-OFDM signal whose band allocation is illustrated in Fig. \ref{fig:structNC2} and is referred to as \textit{structure $2$}. Here, there are two blocks of contiguous active subcarriers of length $c$ which is in range $[3,15]$ and three different interleaved factors $q_1$, $q_2$ and $q_3$, all belonging to the range $[1,8]$. $4$) NC-OFDM signal where the bands are allocated in a random fashion without any structure. In other words, we assume the transmitter flips a coin to decide whether a subcarrier is active or inactive. For this case, the only limitation is the number of active subcarriers which is assumed to be in the range $[4,44]$ for the training and test purposes.

In Section \ref{sec:simulations}, we investigate the performance of the DNN models presented here assuming that they are trained and tested based on one of these $4$ structures. In this way, we are able to see the effect of the band allocations on the performance of the exploiter. Furthermore, the number of signals in the training and test data set for each case is set to $6 M$ and $1.5 M$, respectively.

 To evaluate the performance of the exploiter, we consider the bit error rate (BER) at the receiver when trying to decode the spurious data sent by the exploiter. We assume the receiver knows the original band allocation pattern $\mathcal{S}$, listens to the active subcarriers and decode the corresponding bits. We note that higher errors in exploiter's estimation of $\mathcal{S}$ results in higher BER at the receiver, since the exploiter may be sending on subcarriers that are not considered active at the receiver. Therefore, this BER shows the ability of the exploiter to estimate the transmission parameters and send the desired waveform to the receiver.
\subsection{Utilizing DNN to solve the problem in Example \ref{CAFexample}}\label{Comparison to CAF-based analysis}
We discussed in section \ref{Signal exploitation using cyclostationary analysis} how CAF-based analysis fails to infer parameters of a simple NC-OFDM signal. For a fair comparison, we will consider the same problem in Example \ref{CAFexample} again here, and utilize a DNN to solve it. We also study the effect of the number of hidden layers on the exploiter's performance by considering two structures for the DNN. The first one has a hidden layer with $50$ neurons, and the second one has $3$ hidden layers of $500$, $250$ and $50$ neurons per each layer. The number of neurons at the input layer in both DNNs is set to $768$. The properties of a DNN which estimate the set of parameters $[q,T_u]$ are as follows.
\begin{itemize}
\item {\bf Input:} Samples of the received signal (similar to the previous DNNs).
\item {\bf Output:} Estimated parameters $[\hat{q},\widehat{T_u}]$.
\item{\bf Training:} DNN is trained to minimizes the $L_2$ loss
$\norm{[q,T_u]-[\hat{q},\widehat{T_u}}$ where $\norm{\cdot}$ denotes the L-$2$ norm and $[q,T_u]$ is the true parameters assumed to be known at the exploiter during the training stage. Adam optimizer is used with a learning rate of $0.0001$ to train the model.
\end{itemize}

As another approach to solve the problem, we consider a DNN which classifies between three different classes corresponding to the three cases described in Table \ref{CAFTable}. The specifications for this DNN are:
\begin{itemize} 
\item {\bf Input:} Samples of the received signal (similar to the previous DNNs).
\item {\bf Output:} $\mathbf{\hat{P}}=[\hat{p}_1,\hat{p}_2,\hat{p}_3]$ where $\hat{p}_i$ denotes the probability that the inputs correspond to case $i$, and $\sum_i \hat{p}_i=1$.
\item{\bf Training:} It minimizes the cross entropy loss between $\mathbf{\hat{P}}$ and $\mathbf{P}=[p_1,p_2,p_3]$, i.e., $-\sum_i p_i\log\hat{p}_i$ where $p_i$ represents the true probability that the signal belongs to case $i$. Gradient descent optimizer with a learning rate of $0.0005$ is used for the training.
\end{itemize}
For solving this specific problem, we consider training and test sets of size $5\times10^5$ and $3\times10^5$, respectively, which consist of the signals corresponding to the three transmission cases in Table \ref{CAFTable} at SNR $5$ dB.
\section{Simulation Results and discussion}\label{sec:simulations}
\begin{figure}[ht!]
\vspace{-10pt}
\centering
\includegraphics[width=7.5cm]{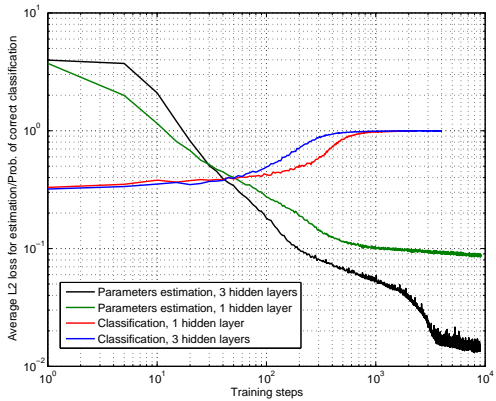}
\caption{Performance of DNN in solving the problem stated in Example \ref{CAFexample} assuming that the same transmission parameters (Table \ref{CAFTable}) are used at SNR $5$ dB. Although Example \ref{CAFexample} shows that CAF is not able to distinguish between the three studied cases, one can use a DNN to either correctly classify these signals with high probability or estimate the original transmission parameters with low error.}
\label{fig:CAFComp}
\end{figure}

We present several numerical examples to evaluate the performance of an exploiter utilizing DNNs for exploiting transmitted signals. We note that the QPSK modulation is used for all the following examples at the transmitter and the exploiter. First, we investigate the performance of the DNNs introduced in Section  \ref{Comparison to CAF-based analysis} for solving the problem (Example \ref{CAFexample}) that CAF failed to solve. This is illustrated in Fig. \ref{fig:CAFComp} where two different cases are considered for each classification and estimation scenarios. Note that the y axis represents correct classification probability for the classification curves, and $L2$-error for the estimation scenarios. Specifically, for the classification scenario, even using one hidden layer enables us to identify the correct class of the signals after around $1000$ training steps with very high probability. In each step, mini batches of size $10^4$ are chosen from the signals in the training set and have been used to train the DNN. If one wishes to estimate the true parameters of the signal, this result shows that it is possible to achieve average L$2$ errors less than $0.1$ using DNN approach. We also observe that higher number of layers results in better performances in both cases.

 Next, we focus on applying deep learning to the signals with band allocation described in Section \ref{DNN Training}. We consider $t_0\in[0,2,4,6,8,10] \mu s$, $N\in[16,32,64]$ and $\Delta f\in[15,20,25,30] KHz$ in the training and test sets.

\begin{figure}[ht!]
\centering
\includegraphics[width=7.5cm]{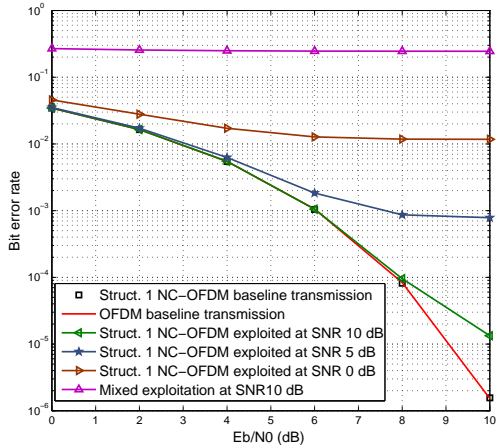}
\caption{The BER at the receiver when the exploiter estimates the transmission parameters and send the signal to the receiver for $3$ cases: OFDM, structure $1$ NC-OFDM (Fig. \ref{fig:structNC1}) and mixed exploitation. Baseline transmission represents the case where all the transmission parameters are known at the receiver.}
\label{fig:BERComp1}
\vspace{-15pt}
\end{figure}

Fig. \ref{fig:BERComp1} demonstrates the performance of an exploiter (using DNN) for signal exploitation in different scenarios where she receives the transmitter's signal at different SNRs, estimate the transmission parameters and tries to send spurious data to the receiver. The bit error rates (BERs) are reported at the receiver for $3$ different cases:  $1)$ For OFDM scenario, the exploiter is able to infer the parameters without error and can achieve the same performance as the baseline transmission (where the parameters are fully known at the receiver). $2)$ For structure $1$ NC-OFDM signal in Fig. \ref{fig:structNC1}, the DNNs are trained and test at $3$ different SNRs $0$, $5$ and $10$ dB. The results show that it is easier to exploit the signals at higher SNRs and lower bit error rates are observed by the receiver for the spurious data over a wide range of $Eb/N0$. However, comparing to the baseline transmission, there is a gap since parameters are estimated with error. $3)$ As the third case, we consider DNNs are trained based on the structure $1$ NC-OFDM signals, and they are tested to infer parameters of OFDM, random NC-OFDM and structure $1$ NC-OFDM signals. We refer to this case as \textit{mixed exploitation}. One can see that this indeed decreases the ability to exploit the signal as one would expect because the DNNs are being tested on signals which have not been trained upon before. 

\begin{figure}[ht!]
\centering
\includegraphics[width=7.5cm]{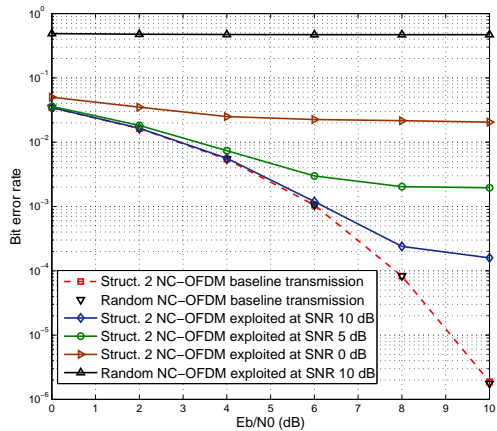}
\caption{The BER described in Fig. \ref{fig:BERComp1} for $2$ more cases: structure $2$ NC-OFDM (Fig. \ref{fig:structNC2}) and random NC-OFDM signal.}
\label{fig:BERComp2}
\vspace{-20pt}
\end{figure}

Similarly, Fig. \ref{fig:BERComp2} illustrates the exploiter's performance for two more cases. First, for structure $2$ NC-OFDM signal in Fig. \ref{fig:structNC2}, one can see that the performance is worse in comparison to the structure $1$ NC-OFDM case in Fig. \ref{fig:structNC1} as the transmitter is using a more complex band structure. Therefore, one can conclude that structure $2$ NC-OFDM signals are harder to be exploited. Second, random NC-OFDM signals are investigated. This result shows that exploiter is unable to exploit such signals because of the high estimation error at the output of the DNNs. One can see that the use of complete random allocation at the transmitter makes the exploiter estimate the parameters with high error, and as a result the corresponding BER at the receiver for the signal that exploiter is sending becomes very close to $0.5$

\begin{figure}[ht!]
\centering
\includegraphics[width=7.5cm]{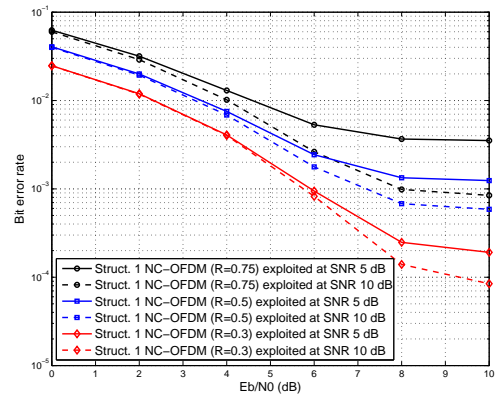}
\caption{Effect of transmission rate in exploiting the signal for structure $1$ NC-OFDM.}
\label{fig:BERRates}
\vspace{-20pt}
\end{figure}

The effect of the transmission rate is studied in Fig. \ref{fig:BERRates}. Specifically, we assume DNNs are trained on structure $1$ NC-OFDM signals of different rates, however, they are tested on three different transmission rates $R\in[0.3,0.5,0.75]$. Furthermore, we fix total number of subcarriers $N=64$ for this example. We note that here rate represents the ratio of the number of active subcarriers to the total number of subcarriers. We observe that as the transmission rate is increased (higher number of active subcarriers) at the transmitter, parameter's estimation error at the output of DNN becomes higher which indeed decreases the signal exploitation ability.

\section{Conclusions}
\label{sec:conclusion}
In this paper, we have studied the robustness of NC-OFDM as compared to OFDM transmissions against DNN-based signal exploitation attacks. In particular, we have considered an exploiter utilizing DNNs to infer the transmit signal waveform so as to be able to send spurious data to the receiver. We have focused on two important signal features (parameters), namely the subcarrier width and the subcarrier allocation pattern, that are necessary for the signal exploitation, and introduced a DNN model for the exploiter to infer these parameters from samples of the signal that she receives. Numerical examples have shown that the structure of the NC-OFDM band (subcarrier occupancy pattern) plays an important role in the parameter estimation performance at the exploiter. Specifically, as the transmitter chooses the NC-OFDM subcarrier occupancy pattern in a more random fashion, it becomes harder for the exploiter to train a DNN model to estimate the parameters and exploit the signal. This suggests that a transmitter should avoid using structured band allocations in order to protect the system against DNN-based signal exploitation attacks. 
\bibliographystyle{IEEEtran}
\bibliography{references}

\end{document}